\def\hi{\ion{H}{1}}
\def\hii{\ion{H}{2}}

\newcommand{\beq}{\begin{equation}}
\newcommand{\eeq}{\end{equation}}
\documentclass[12pt,preprint]{aastex}
\shortauthors{GARNETT}
\shorttitle{EFFECTIVE YIELDS AND METAL LOSS IN GALAXIES}
\received{ }
\revised{____________}
\accepted{ }
\slugcomment{Accepted by the {\it Astrophysical Journal} 19 Aug 2002}
\begin{document}
\title{The Luminosity-Metallicity Relation, Effective Yields, and 
Metal Loss in Spiral and Irregular Galaxies}
\author{Donald R. Garnett} 
\affil{Steward Observatory, University of Arizona, 933 N. Cherry Ave., 
Tucson, AZ  85721 \\ E-mail: dgarnett@as.arizona.edu}

\begin{abstract}

I present results on the correlation between galaxy mass, luminosity, and 
metallicity for a sample of spiral and irregular galaxies having well-measured 
abundance profiles, distances, and rotation speeds. Additional data for low 
surface brightness galaxies from the literature are also included for comparison. 
These data are combined to study the metallicity-luminosity and 
metallicity-rotation speed correlations for spiral and irregular galaxies. 

The metallicity luminosity correlation shows its familiar form for these
galaxies, a roughly uniform change in the average present-day O/H abundance 
of about a factor 100 over 11 magnitudes in B luminosity. 
However, the O/H - $V_{rot}$ relation shows a change
in slope at a rotation speed of about 125 km s$^{-1}$. At faster $V_{rot}$,
there appears to be no relation between average metallicity and rotation
speed. At lower $V_{rot}$, the metallicity correlates with rotation speed.
This change in behavior could be the result of increasing loss of metals
from the smaller galaxies in supernova-driven winds. This idea is tested 
by looking at the variation in effective yield, derived from observed
abundances and gas fractions assuming closed box chemical evolution. The 
effective yields derived for spiral and irregular galaxies increase by a 
factor of 10-20 from $V_{rot}$ $\approx$ 5 km s$^{-1}$ to $V_{rot}$ $\approx$
300 km s$^{-1}$, asympotically increasing to approximately constant $y_{eff}$ 
for $V_{rot}$ $\ga$ 150 km s$^{-1}$. The trend suggests that galaxies with
$V_{rot}$ $\la$ 100-150 km s$^{-1}$ may lose a large fraction of their
SN ejecta, while galaxies above this value tend to retain metals.


\end{abstract}

\keywords{Galaxies: abundances --- galaxies: evolution --- galaxies: spiral
--- galaxies: irregular}

\section{Introduction}

The strong correlation between galaxy metallicity $Z$ and galaxy luminosity $L$
is one of the more significant phenomenological results in galaxy chemical 
evolution studies. The basic correlation of O/H vs. $L$ was demonstrated for 
irregular galaxies by \cite{lrspt79} and confirmed by \cite{skh89}. \cite{gs87} 
extended this correlation to include spiral galaxies, demonstrating that a 
correlation between metallicity and luminosity extends over a factor of 100
in metallicity and 11 magnitudes in blue luminosity $M_B$. Parallel studies 
of elliptical galaxies \citep{faber73,bh91} showed a similar metallicity-luminosity 
correlation. \cite{zkh94} noted that both ellipticals and star-forming galaxies 
exhibited similar metallicity luminosity correlations, despite the different 
measurement techniques. This suggests that similar phenomena govern the 
metallicity-luminosity relationship in spiral/irregular and elliptical galaxies.

It is not clear that the primary correlation in this case is between
metallicity and luminosity. For disk galaxies, luminosity, rotation
speed (or total mass), surface brightness, and Hubble type are all 
correlated to some degree (Roberts 1994; de Jong \& Lacey 2000), so 
it is not certain which is the main driver of the correlation. Rotation 
speed ($V_{rot}$), taken as the maximum measured speed or the speed on 
the flat part of the rotation curve, or $V^2_{rot}R$, where $R$ is the 
radius, may be preferable measures of galaxy mass, but \cite{zkh94} 
and \cite{gar97} found that the metallicity-$V_{rot}$ correlation was 
not noticeably better than the metallicity-luminosity correlation, at 
least for spiral galaxies.

The origin of this correlation is open to debate. A metallicity-luminosity
correlation can arise if smaller galaxies have larger gas fractions than
larger galaxies, which is seen statistically in the local universe 
\citep{rh94,mdb97,bdj00}. This situation can arise if small galaxies either 
evolve more slowly (lower star formation rate per unit mass) or are younger 
on average than more massive systems, and have thus simply processed a smaller 
fraction of their gas into stars. The relatively blue colors of low-mass 
irregular galaxies indicates that this 
must be true at some level.  On the other hand, it has been popular
recently to ascribe the correlation to the effects of selective loss of
heavy elements from galaxies in supernova-driven outflows. Low-mass galaxies
are expected to lose a larger fraction of SN ejecta than more massive systems
because of their smaller gravitational potentials. However, the loss of gas
from galaxies depends not just on gravitation potential, but also on the
vertical structure of the ISM and details of radiative losses, among other
things. Numerical modeling of outflows by \cite{mlf99} showed that it is
extremely difficult to `blow away' the ISM of a gas-rich dwarf galaxy via
starburst superwinds, although selective loss of metals may be relatively
easy. However, a generalized description of the conditions for loss of 
metals from galaxies is not yet available.

The question of loss of metals from galaxies is of high interest because
of the apparently ubiquitous presence of metals in the intracluster medium
(ICM) and the intergalactic medium (IGM; Ellison et al. 2000). 
The source of these metals, whether from galactic outflows, tidal/ram
pressure stripping in dense environments, or pre-galactic stars, is debated.
It is therefore worthwhile to look for evidence that galaxies are actually
losing metals to the IGM. I begin here by examining the $L$-$Z$ correlation
in a different way. Figure 1(a) shows the familiar O/H - $M_B$ correlation
for spiral and irregular galaxies, taking the value of O/H at the disk
half-light radius $R_{eff}$ as a surrogate for the average ISM abundance
in spirals; the data used is discussed below. 

B-band magnitudes are often considered to be a less-than-ideal surrogate for 
galaxy mass, however. The blue light from galaxies can be strongly affected 
by interstellar extinction, and variations in recent star formation history 
can make the relation between stellar mass and blue luminosity uncertain, 
especially for bursting dwarf galaxies. Therefore, in Figure 1(b) I plot O/H 
versus rotation speed $V_{rot}$ (obtained from rotation curves) for the same 
set of galaxies; data for this plot are listed in Tables 1-3. This panel 
shows that, interestingly, the correlation between 
O/H and $V_{rot}$ does not increase steadily, but rather turns over for rotation
speeds greater than 125 km s$^{-1}$. Above this value, the data indicate that 
the mean metallicity for the most massive spirals is essentially constant. 
(To a large degree, this behavior becomes apparent because we have changed
from a logarithmic scale [B magnitude] to a linear scale [$V_{rot}$]). This 
result suggests that there is a velocity/mass threshold below which the 
metallicity evolution of a galaxy varies according to mass. 

\section{Gas Flows and Effective Yields}

Gas flows, both infall and outflow, affect the metallicity evolution of a 
galaxy in measurable ways. \cite{edm90} has examined the effects of such
flows in the context of the closed box chemical evolution model. Under
the assumptions of no gas flows and instantaneous recycling of stellar
ejecta, metallicity evolution in the closed box model is a simple function 
of the gas fraction $\mu_{gas}$ and the true yield $y_Z$ given by the 
familiar expression 

\begin{equation}
Z = y_Z~ln(\mu^{-1}).
\end{equation}
One can write a corresponding expression for any individual element by 
replacing $Z$ and $y_Z$ with the appropriate quantities for individual 
elements.

By inverting equation (1), one can define the effective yield $y_{eff}$:

\begin{equation}
y_{eff} = {Z(obs) \over ln(\mu^{-1})}.
\end{equation}
The effective yield is thus a quantity that can be measured from the
observed metallicity and gas fraction. 

We assume for the purposes of this work that the true yield is essentially
constant. The true yield is a function of stellar nucleosynthesis. Massive 
stars ($M$ $>$ 10 $M_{\odot}$) account for most of the bulk nucleosynthesis 
of elements heavier than helium, and oxygen comprises the largest fraction 
of this, 45-60\% of $Z$ by mass. Nucleosynthesis is somewhat dependent on
metallicity, although the direct effects on bulk nucleosynthesis in stellar
cores are relatively small (Woosley \& Weaver 1995; see Fig. 2.10 in Matteucci
2001). The largest influence of metallicity on nucleosynthesis in massive stars 
is the effect of radiatively-accelerated mass loss in stellar winds \citep{m92}. 
Stellar winds in massive stars can remove He and C that would otherwise be 
converted to O, thus 
increasing the yields of He and C and reducing the yield of O. These effects
are most serious for stars more massive than 25 $M_{\odot}$ at metallicities 
of roughly solar and above. Integrated over a \cite{sal55} IMF, the net
yield for O in this case declines by no more than a factor of two between
$Z$ = 0.002 and $Z$ = 0.02 \citep{m92}. For the no-mass-loss models of
\cite{ww95}, a weak increase in the bulk yield is seen over the same range
of $Z$.

\cite{edm90} demonstrated in a series of theorems that one could derive
a few general properties of chemical evolution in the case of outflows
and inflows:

(1) In the case of pure outflows, the effective yield is always smaller
than in the case of the simple model.

(2) In the case of inflow of unenriched gas, the effective yield is always
smaller than that in the simple model whether or not there is an outflow.

(3) If the inflow is enriched, the metallicity of the infalling gas must
be comparable to the metallicity of the system to significantly alter
inference (2).

(4) If outflow is greater than infall, the effective yield is smaller 
than that of the simple model as long as the infalling gas has metallicity
less than that of the system.

Figure 2, adapted from Figure 1 of \cite{edm90}, illustrates the situation
for the case of outflows or unenriched inflows. The simple model is a 
straight line in the diagram whose slope is the yield. There is a region
of the diagram with $Z$ $>$ $Z$(simple model) which is forbidden to systems
with outflows or unenriched inflows. In these cases the effective yield 
is always smaller than the true yield. 

Thus, in the context of the simple chemical evolution model, it is possible
to investigate how effective gas flows are in the metallicity evolution 
of galaxies. If the $Z$-$L$ correlation is simply the result of variation
of gas fraction along the mass sequence, then the effective yields derived
for the galaxy sample should be relatively constant (modulo the possibility
of a small metallicity dependence of the true yields). If inflow and outflow
of gas are important, and if they depend on the galaxy potential, then one 
might expect to see a systematic variation of effective yield with galaxy mass. 

\cite{mc83} looked at the variation of metallicity versus gas fraction
in a sample of dwarf irregular and blue compact galaxies, and noted that 
the effective yields derived for these galaxies were quite low compared
to the expected yields. They suggested that outflows of gas could account
for the apparent low observed values of $Z$ with respect to gas content.
However, they derived total baryonic masses from rotation curves, whereas
it is likely that the dynamical mass is dominated by non-baryonic dark
matter. Thus, it is probable that \cite{mc83} underestimated the baryonic 
gas fraction. Here I will use the measured galaxy luminosity to estimate 
the stellar mass.

Similarly, \cite{vce92} studied effective yields within a variety of spiral
galaxies, finding evidence for a variation of $y_{eff}$ with metallicity. 
Since then, a great deal of improved data on abundances (for example, the 
surveys of Zaritsky, Kennicutt \& Huchra 1994, van Zee et al. 1998, and 
Ferguson et al. 1998) and CO measurements (for example, the FCRAO survey 
of CO in galaxies, Young et al. 1995) have become available for both spiral 
and irregular galaxies. I will use these new measurements to conduct an 
expanded investigation of effective yields in galaxies. The following 
sections will describe the data used and the results derived for global 
effective yields.

\section{The Sample and Data}

The sample of 31 spiral and 13 irregular galaxies presented here was chosen 
to have spectroscopic data on \hii\ region abundances covering a large 
fraction of the galaxy, together with atomic and molecular gas masses from 
large beam measurements or wide-area maps. Spiral galaxies were chosen
primarily from the spectroscopic surveys of \cite{zkh94,vz98,fgw98},
although a number of galaxies were chosen from various individual studies,
cited in Table 1.
The irregular galaxies were chosen largely from the nearby sample listed
in \cite{mateo98}; the \hii\ region-like blue compact dwarf galaxies are 
generally excluded because their intense star formation activity makes it 
difficult to estimate the underlying stellar mass density from the blue 
luminosity. Tables 1-3 list the galaxies in the sample, their structural 
properties, and the abundance and gas data adopted. The criteria for 
selection of data is described below.

\subsection{Abundances}

I use radial O/H distributions obtained from direct spectroscopic studies
of \hii\ regions for consistency. Measurements of O/H derived from narrow-band 
imaging studies exist for a number of galaxies (mainly barred spirals),
but such measurements typically show a large amount of point-to-point
scatter which could reflect real O/H variations but could also be due to
variations in excitation, so it is not completely clear if these data
can be compared to direct spectroscopic measurements in a consistent way.
For this study I use the value of O/H at the B-band effective (half-light)
radius of the disk, $R_{eff}$, as a surrogate for the mean metallicity of
the young component. $R_{eff}$ = 1.685 times the exponential scale length
$R_d$ \citep{dvp78}, easily derived from disk photometric properties.
O/H($R_{eff}$) is determined by interpolating the derived O/H gradient.
The accuracy of this determination depends on the number of \hii\ regions
measured and the radial fraction of the disk sampled by the observations
\citep{zkh94}. To ensure relatively precise values of O/H($R_{eff}$) while
retaining a useful number of galaxies to study, I selected spiral galaxies
with at least five \hii\ region measurements covering a radial range
$\Delta R/R_{25}$ $>$ 0.5, where $R_{25}$ is the radius at which the surface
brightness falls to 25 magnitudes per square arcsecond and $\Delta R$ is
the difference in radius between the innermost and outermost observed \hii\
regions. For comparison, I include results based on using the value of O/H 
at one disk scale length, O/H($R_d$); this change has little effect
on the overall results.

A small fraction of the interstellar oxygen in the neutral/ionized gas 
is incorporated into metal oxide and silicate grains. This fraction is 
expected to be no larger than 15-20\%, based on the maximum amount that 
can be used if all of the interstellar Mg, Si, and Fe were depleted into 
such grains. Ices are not a significant contribution to depletion of O 
in the diffuse ISM \citep{mathis90}. Observations of O/H from interstellar 
absorption lines in the local ISM suggest that there is little variation 
in the fraction of O that could be in grains \citep{mjc98}. Whether the
fraction of the heavy element abundance that is locked up into grains
varies with metallicity is poorly known, although the dust-to-gas ratio
does appear to vary with metallicity. Nevertheless, the fraction of O that 
is not counted because it is in grains can be argued to be insignificant
compared to other contributions to the uncertainties of the derived 
effective yields.

\subsection{Atomic Gas}

Atomic gas typically accounts for the majority of the gas content of
a spiral or irregular galaxy. The atomic component tends to be very
extended compared to the stellar disk in spirals and irregulars, as
much as 3-4 times the photometric radius $R_{25}$ in spirals \citep{bvw94} 
and even more in dwarf irregulars. Thus, for nearby galaxies with large
angular sizes such as M101 the atomic gas angular extent can be 1$^\circ$
or more, necessitating large-scale mapping to determine the total gas
content. 

For this study I use the total \hi\ content derived from either single
dish measurements or fully sampled maps. I have tried to avoid using
aperture synthesis \hi\ maps because aperture synthesis measurements tend 
to lose diffuse flux on large scales due to missing short antenna spacings, 
and because the fields covered are often not large enough to include 
the outer regions of the largest nearby galaxies. Fortunately, many of 
the nearby galaxies have either large-beam single-dish measurements or 
fully-sampled 21 cm maps over large enough scales to sample all of the 
atomic gas on kiloparsec size scales, adequate for studies of the total 
mass and bulk surface density distribution. The atomic gas masses adopted 
for the sample galaxies, multiplied by a factor 1.33 to account for helium, 
are listed in Tables 2 and 3, along with the references for the 21cm line 
measurements. In cases where multiple measurements exist, I have used the 
curve of growth of 21cm flux versus beam (or map) size to estimate the 
total \hi\ mass and its uncertainty. In most cases the uncertainty in the 
total \hi\ mass is about 10-15\%. 

\subsection{Molecular Gas}

The molecular gas component is the environment for star formation, and 
can also contribute significantly to the gas surface density in the inner 
disks of spirals. Unfortunately, cold H$_2$ is very difficult to observe 
directly; the CO molecule is used as a tracer of the molecular component, 
with considerable debate over the conversion from CO flux to H$_2$ column 
density and its dependence on metallicity and enviroment. Moreover, 
compared to 21cm measurements, relatively few galaxies have been fully 
mapped in CO. The CO flux falls rapidly with radius outside the inner
few kiloparsecs of many spirals, and is vanishingly faint in irregular
galaxies.

I have collected CO measurements from the literature for the sample 
galaxies, and have used fully sampled maps to estimate the H$_2$ mass
where possible; otherwise, most of the adopted integrated fluxes have
been taken from the FCRAO survey \citep{young95}. The same cautions
on using aperture synthesis observations for H I masses applies to
CO measurements as well. Of the 31 spirals listed in Table 1, only 23 
have CO measurements suitable for estimating the mass of molecular gas.

To estimate M(H$_2$) for the galaxies in my sample, I have converted 
the measured CO fluxes from the literature adopting a constant 
N(H$_2$) - I(CO) conversion factor of 
3$\times$10$^{20}$ cm$^{-2}$ (K km s$^{-1}$)$^{-1}$ 
\citep{wilson95}, which includes the correction for helium. I adopt
this approach for several reasons:

(1) In spirals, CO is detected largely in regions with metallicities
within a factor of two or so of the solar value. If there is a systematic
variation in N(H$_2$)/I(CO) with metallicity, it is probably modest over 
this range of O/H. 

(2) Assuming this conversion, the fraction of molecular gas is typically
less than 50\% in massive spirals, and much less in low-mass galaxies
(Figure 3). Thus, the contribution to the error in the total gas mass
due to the uncertain metallicity correction is probably small compared
to the uncertainty in the standard conversion factor. 

(3) In metal-poor irregular galaxies, the CO fluxes are so small that
the derived H$_2$ masses are negligible compared to the \hi, even if
one assumes a strong metallicity dependence of 
N(H$_2$)/I(CO) 
as in
\cite{israel97}. An illustration of this is given by the case of NGC 
4214, a highly active star-forming irregular galaxy with 12 + log(O/H)
= 8.2 \citep{ks96}. \cite{ty85} derived M(H$_2$) = 4$\times$10$^6$ M$_\odot$
assuming 
N(H$_2$)/I(CO) = 4$\times$10$^{20}$, while \cite{thron88} estimated
M(H$_2$) $\approx$ 10$^8$ M$_\odot$ assuming
N(H$_2$)/I(CO) = 2$\times$10$^{21}$. From \cite{israel97} we derive
N(H$_2$)/I(CO) = 6.3$\times$10$^{21}$ for 12 + log O/H = 8.2, which
would increase the above values for M(H$_2$) to 6.3$\times$10$^7$ 
M$_\odot$ and 3$\times$10$^8$ M$_\odot$, respectively. This can be 
compared to the atomic gas mass of 2.4$\times$10$^9$ M$_\odot$
\citep{hgr82}. Thus, even with the strong metallicity dependence 
of \cite{israel97} the molecular gas in dwarf irregulars is only a
small fraction of the total gas mass.

\subsection{Stellar Component}

Masses for the stellar component of galaxies may be even more uncertain
than the molecular gas content. Dynamical masses can not be used to 
estimate the stellar component, so we must rely on converting the 
measured luminosity to mass via an estimated mass-to-light ratio, $M/L$. 
$K$-band photometry is less sensitive to variations in $M/L$ and dust 
attenuation than optical photometry, but wide-field IR imaging, or even 
$R$ or $I$-band imaging, is generally unavailable for the nearby galaxies 
with large angular sizes. 

For this study I used the integrated $B$-band magnitudes and $B-V$ colors
from the RC3 \citep{rc3} to estimate masses for the stellar components.
To see the effects of uncertainties in $M/L$ on the results I derived
the stellar component for the spirals two different ways: (1) First,
I used the relations for $M/L$ versus color from \cite{bdj01}, who
showed that $M/L$ is quite well correlated with color for a variety
of optical bandpasses based on the results of population synthesis
models. (2) Second, I assumed a constant $M/L$ = 2 for all spiral
galaxies. This is roughly an average $M/L$ for disks derived from
photometric and kinematic analyses, e.g., \cite{wevers86}. For the
irregulars, I assumed a constant $B$-band $M/L$ = 1. This number is
fairly uncertain; the population synthesis models become degenerate
at low metallicities, and the bursty star formation histories of 
irregulars makes comparison with simple stellar population models
difficult. As we shall see, a factor two uncertainty in $M/L$ will
not significantly affect the trends I derive. 

\subsection{Gas Fractions}

As derived here, the gas fraction $\mu_{gas}$ is simply the total
gas mass divided by the sum of the gas mass and the mass in stars.
It is fair to ask whether the reservoir of gas that lies external
to the stellar disk should be counted toward the total, on the 
grounds that such gas may never have participated in star formation,
and so may not have abundances that can be smoothly extrapolated
from the distribution in the inner disk. On the other hand, to 
estimate $y_{eff}$ we need to know the amount of gas that $could$ 
participate in star formation, and since viscous processes can mix 
gas from the outer disk to the inner regions, the gas in the outer 
disk may be very relevant to the total. Since irregular galaxies 
have a larger fraction of gas outside the stellar disk than spirals, 
a discontinuous drop in the metallicity of the outer gas could 
introduce a bias in comparing effective yields. 

One can roughly estimate the relative effect by assuming that the
gas outside the stellar disk has a much lower metallicity. I will
assume for illustration purposes that 90\% of the gas in an irregular 
lies outside the region where stars are found. I will also assume 
that this fraction of the gas has O/H = 1$\times$10$^{-6}$ 
(approximately 0.001 times solar), while the remaining gas has 
O/H = 2$\times$10$^{-5}$ (similar to the value in I~Zw~18 [Skillman
\& Kennicutt 1993]). For the example spiral galaxy, I use the 
result of Broeils \& van Woerden (1994) that the average ratio
of \hi\ diameter to stellar disk diameter D(\hi)/D(disk) is 
1.8$\pm$0.6. If we further assume that the gas distribution is flat, 
the average D(\hi)/D(disk) implies 76\% of the {\it atomic} gas is 
outside the stellar disk. (This will likely overestimate the amount 
in the outer regions, given that \hi\ column densities tend to drop 
outside the stellar disk.) On the other hand, the molecular gas 
component is located mostly within the stellar disk radius. 
A reasonable assumption for illustration purposes is that 60\% of
the gas in a typical spiral is within the stellar disk, while the
rest is outside.
I will assume that the average O/H of the 
gas within the stellar disk radius is solar (8$\times$10$^{-4}$), 
while outside the disk O/H = 1$\times$10$^{-6}$. 
(The widespread observation of metals in the IGM at high redshifts 
\citep{essp00} suggests that the metallicity in the outer regions
of galaxies is unlikely to be zero.) 

With these assumptions, the average O/H for the irregular galaxy 
would be 2.9$\times$10$^{-6}$, seven times smaller than the value 
within the stellar disk, while for the spiral the average O/H 
would be 4.8$\times$10$^{-4}$, 60\% of the value within the 
stellar disk radius. Thus, in these cases one would underestimate
$y_{eff}$ in the irregular galaxy by a factor seven, and $y_{eff}$
in the spiral galaxy by a factor 1.7, leading to a relative bias
of a factor four in the difference in $y_{eff}$ between the irregular 
and the spiral. This rough estimate of the possible bias should be 
kept in mind during the discussion of the results shown in Figure 4. 

Table 4 shows the derived gas fractions and effective yields for
oxygen for the spiral and irregular galaxies that appear in Figures 
4-6. The oxygen effective yields listed in the table are those 
computed assuming the average oxygen abundance is the value at 
$R_{eff}$. These are the values plotted in Figure 4(a).

\section{Results}

Figure 4 shows the basic results from this analysis, the values of 
$y_{eff}$ for oxygen derived for each galaxy plotted versus $V_{rot}$. 
Fig. 4(a) shows $y_{eff}$ based on O/H at $R_{eff}$ and stellar masses 
assuming color-based $M/L$. Fig. 4(b) shows the results if constant 
$M/L$ = 2 is assumed for the stellar component, and Fig. 4(c) shows 
what happens if we use O/H at $R_d$ as the average. Note that
the values of $y_{eff}$ for the irregular galaxies do not change in
the three panels, as I use the same assumptions to estimate $y_{eff}$.
The trend in $y_{eff}$ vs. $V_{rot}$ is very similar in all three
panels, showing that the calculation of $y_{eff}$ is fairly robust
to changes in the assumptions; these affect mainly the scale of 
$y_{eff}$ in the spirals. The scatter in $y_{eff}$ at a given $V_{rot}$
is fairly large, about $\pm$0.2 dex, reflecting mainly the uncertainties 
in disk $M/L$ and $M(H_2)$.

Despite the uncertainties, Figure 4 shows a systematic increase
in $y_{eff}$ by a factor of 10 to 20 as $V_{rot}$ increases from a few
km s$^{-1}$ up to 300 km s$^{-1}$. At this point, we return to the
discussion at the end of Section 3.5 and note that if the large gas
reservoirs in the outer parts of spirals and irregulars are more 
metal-poor than expected from extrapolation of the inner disk abundances,
then $y_{eff}$ could be biased toward lower values
by a factor of 4-5 relative to the spirals. Nevertheless, Figure 4 shows 
that the smallest irregular galaxies are clearly underabundant relative 
to spirals, if the true yield does not vary much with metallicity.
Note that the dashed line corresponding to the solar value for $Z(O)$,
the solar oxygen mass fraction, is for reference only. 

It is also apparent from Fig. 4 that log($y_{eff}$) does not increase 
uniformly with $V_{rot}$. Most of the increase occurs for $V_{rot}$
$<$ 100 km s$^{-1}$. For $V_{rot}$ $\ge$ 200 km s$^{-1}$, $y_{eff}$
appears to constant (within the scatter). Thus, $y_{eff}$(O) appears
to asymptotically approach a constant for galaxies with $V_{rot}$
$>$ 150-200 km s$^{-1}$. There is an unfortunate gap in data for
galaxies with 140 km s$^{-1}$ $<$ $V_{rot}$ $<$ 200 km s$^{-1}$, so
where this approach toward constant $y_{eff}$ occurs is not as clear
as it could be. The reason for this gap is not clear, but it seems
be due mainly to one or another missing piece of information (such
as \hi\ or H$_2$ data), rather than to sample bias. 

The interesting question for purposes of the discussion here is at
what point we can say that $y_{eff}$ is essentially constant with 
$V_{rot}$? It is fair to say that this is true for $V_{rot}$ $>$
200 km s$^{-1}$. The gap for rotation speeds between 140 and 200 
km $^{-1}$ makes it difficult to determine if there is a clear 
demarcation. Another factor is the potential bias discussed in Section 
3.5. If such a bias in determining $y_{eff}$ exists, then the
turnover point could be at significant lower $V_{rot}$, perhaps as
low as 60-80 km s$^{-1}$. This argues that measurements of the 
metallicity in the outer gas reservoirs of both irregulars and spirals 
should be a high priority for future observations. Another major source 
of uncertainty is $M/L$ for the stellar component; if $M/L$ for the
stellar component in the irregulars is significantly smaller, we 
would derive larger values of $y_{eff}$ for those galaxies and the 
turnover in $y_{eff}$ could be driven down to smaller $V_{rot}$. 
Better estimates of $M/L$ for stellar populations should also be
a high priority for the future.

It is also of interest to see if other types of star-forming galaxies
follow the same trends seen in Figure 4. For example, Figure 5 shows
effective yields derived for low surface brightness (LSB) galaxies
from the sample of \cite{vdhoek00}, displayed as open circles. In
this case the gas fractions and yields have been derived assuming a 
constant $M/L$ = 1 for the LSB galaxies, rather than using the values
assumed by \cite{vdhoek00} based on maximal disk models for the 
rotation curve (which probably overestimate the $M/L$ for the disk).
In most cases, my assumption for $M/L$ leads to larger gas fractions
than obtained by \cite{vdhoek00}, and thus to larger values of $y_{eff}$.

Figure 5 shows that the smaller LSB galaxies have derived $y_{eff}$
values similar to those of the irregulars, but that the more massive
LSB galaxies appear to fall below their high surface brightness 
counterparts. This is curious, as a plot of $y_{eff}$ vs. average
surface brightness (not shown here) shows the LSBs to fall among 
irregular galaxies with similar average surface brightness. One concern 
is that the LSB galaxies with the largest
$V_{rot}$ values tend also to have small estimated inclinations, and
the LSB galaxies studied by \cite{vdhoek00} have sizes only a few times 
larger than the beam size of the \hi\ measurements. Relatively small 
errors in the inclination could lead to a large error in $V_{rot}$, 
especially if the gas has a different inclination than the stellar
component, as is often seen for nearby irregular galaxies. However,
if the rotation speeds are accurate, it would follow that LSB spirals
tend to have smaller $y_{eff}$ than HSB galaxies with comparable 
$V_{rot}$. If this is true, then it can not be said that LSB galaxies
are simply slowly-evolving counterparts to HSB spirals.

\cite{sksz96} presented evidence that Virgo cluster spirals with
greatly truncated \hi\ disks have higher ISM abundances than field
spirals with comparable $M_B$, $V_{rot}$, and Hubble type. 
\cite{sksz96} noted that the apparent enhancement of O/H is largest
for those Virgo spirals with the most disrupted gas disks, which are 
presumed to have passed through the center of the cluster; meanwhile 
spirals with relatively normal \hi\ disks have O/H comparable to 
similar field spirals. They argued that in the cluster environment 
spiral galaxies have their gas reservoirs stripped, reducing metal-poor 
infall and leading to an apparent enhancement of the metallicity in the 
cluster spirals.  This idea can be tested by looking at $y_{eff}$. 
In the simple chemical evolution model with metal-poor infall (but no 
outflow), truncation of the infall would result in the galaxy evolving 
more like a closed box model. The effective yield in this case would 
be higher than for a galaxy that is experiencing infall (Fig. 2). Thus 
we should expect the stripped Virgo spirals to have larger effective 
yields than those for the field spiral sample. Figure 6 shows this 
comparison, where the stripped Virgo spirals from \cite{sksz96} are 
plotted as unfilled squares. No systematic offset is seen between the 
Virgo spirals and the field spirals. It is possible that the considerable 
uncertainties in $y_{eff}$ preclude the detection of such an offset, and 
the Virgo sample may be too small to show an offset statistically, but 
the straightforward interpretation of Figure 6 is that the high abundances
in stripped Virgo spirals is not a result of truncated cosmological infall.

\section{Discussion}

As noted in Section 2, either outflows or metal-poor inflows can reduce
the effective yield. Then how do we interpret the results shown in Figure 
4? The trend of decreasing $y_{eff}$ with decreasing $V_{rot}$ and galaxy
luminosity suggests increasing importance of SN-driven outflows in the
smallest galaxies, as suggested by \cite{ds86}, and we tend to favor this
model as the simplest interpretation. The smooth increase of $y_{eff}$ 
with increasing $V_{rot}$ would then imply that the fraction of material 
lost is a simple function of the galaxy potential. Flattening of the trend 
for $V_{rot}$ $>$ 100-150 km s$^{-1}$ would indicate that such massive 
galaxies essentially retain all of the metals produced by stars in those
galaxies; gas may be driven into the halo by SN-driven outflows, but the 
ejected gas eventually rains back down onto the galaxy to enrich the disk.
\cite{martin99}, using a different argument, suggested that escape of hot 
gas in galactic winds occurs for galaxies with $V_{rot}$ $<$ 130 km s$^{-1}$,
similar to what I find here.
Alternatively, it would be necessary to posit a conspiracy between inflow
and outflow that keeps the massive spirals at roughly the same $y_{eff}$
regardless of mass.

In a strict sense, it is not possible to rule out pure inflow of metal-poor 
gas as a cause of the trend in Figure 4 from this data alone. The interpretation
of Figure 4 in this case would be that dwarf galaxies have accreted a 
larger fraction of outside gas at relatively late times compared to 
spirals. Evidence for inflow of gas onto dwarf galaxies is much more 
scarce than evidence for outflows, although \cite{ks95} and \cite{tbh97}
argue that NGC 5253 may be accreting gas that is fueling the current
starburst, and \cite{wm98} show that IC 10 has extended plumes of \hi\
that they interpret as gas infalling onto the galaxy. However, to explain
the low effective yields for the dwarf galaxies would require as much
as 80-90\% of their gas to have been accreted at late times without much 
star formation and subsequent metal enrichment. This was demonstrated
by \cite{ke99} for galaxies for a wide range in the ratio of accretion 
rate to star formation rate. Only models in which gas was accreted much
faster than the star formation rate acheived low effective yields; galaxies
with slow accretion (and no outflows) tend to have effective yields that 
approached the true yield with time. Local Group dwarf irregular galaxies 
show a large range of star formation histories, but they are consistent 
with a roughly constant, if low, star formation rate over the past 10 Gyr 
\citep{mateo98}, so unless they have all accreted large amounts of gas in
the recent past, it would be difficult to explain their low effective 
yields with pure infall. 

Stripping of gas can also reduce $y_{eff}$ by decreasing the gas fraction
at a fixed metallicity. This could be particularly important for low-mass
satellites of large galaxies. Stripping could indeed lead to a trend like
that seen in Figure 4, since dwarf galaxies are more likely to be stripped
of gas than the more massive spirals. This could certainly be relevant to
galaxies like the Magellanic Clouds, where evidence for tidal stripping
is strong. Moreover, the detection of hot gas in even small groups of
galaxies \citep{mdmb93} makes it more likely that ram pressure can play 
a role even in relatively low-density galaxy environments. \cite{bc02}
cite Holmberg II as an example of possible gas stripping from a dwarf
galaxy despite its location on the edge of the M81 group, and IC 10 could 
also fall into this picture. On the other hand, Sextans A, which is on
the edge of the Local Group, shows no evidence 
for a disturbed gas disk \citep{wh02}. Sorting out the origin of 
extended gas structures in dwarf irregulars will require deep \hi\ 
mapping of many more galaxies. In the meantime, the dwarf irregulars 
included in this sample are relatively isolated, so we might expect that
gas stripping is less likely to be important in determining the effective 
yields. Furthermore, stripping is much less efficient in reducing the
effective yield than direct loss of metals: very small galaxies like 
Leo A would have to have some 80-90\% of their gas stripped to account 
for their low effective yields. 

If we accept the premise that the variation of $y_{eff}$ is the result
of increasing importance of outflows of metals in low-mass galaxies, then 
the rotation speed tells us which galaxies are likely to be enriching
the IGM. We can then discuss the fates of outflows observed in various
starburst galaxies. For example, we can infer that low-mass starburst
galaxies such as I Zw 18 ($V_{rot}$ $\approx$ 30 km s$^{-1}$) and 
NGC 1569 ($V_{rot}$ $\approx$ 40 km s$^{-1}$) probably lose a large
fraction of the metals produced by their stars into the IGM. On the 
other hand, in more massive spirals with outflows, such as NGC 253 
($V_{rot}$ $\approx$ 210 km s$^{-1}$) and NGC 4631 ($V_{rot}$ $\approx$ 
150 km s$^{-1}$), metals ejected into their halos probably remain bound, 
to fall back onto the disk in a fountain. M82 presents a more ambiguous
case because of its peculiar rotation curve: the galaxy has a maximum
rotation speed of about 200 km s$^{-1}$, but this declines with distance 
from the nucleus \citep{sofue97}. Another complication is its location
in the dense M81 group environment, since gas ejected into the halo 
can suffer tidal stripping. 

If the fraction of metals that is lost by a galaxy is known, it is 
possible to estimate the contribution to IGM enrichment by galaxies
of a given $L$ or $V_{rot}$. This is important to understanding the
element abundance pattern in the IGM, as dwarf galaxies have a different
element abundance pattern than spirals. Given the present uncertainties
in data for effective yields, it is not yet possible to say with great
precision what range of galaxies contribute most to IGM enrichment, 
but we can make a crude estimate of the relative contribution to IGM
enrichment as a function of $M_B$.

I approach this by assuming a Schechter luminosity function of the form

\begin{equation}
\phi(L) = (\phi^*/L^*) (L/L^*)^{\alpha} exp(-L/L^*),
\end{equation}
where $L$ is the B-band luminosity, and I adopt $\alpha$ = $-$1.2,
log $L^*$ = 10.16 ($M_B$ = $-$20.2) and $\phi^*$ = 1.0$\times$10$^{-2}$
based on the field galaxy studies of \cite{ellis96}, \cite{blanton01},
and \cite{folkes99}, scaled to a Hubble constant of 75 km s$^{-1}$ 
Mpc$^{-1}$. The actual normalization of $\phi^*$ is not important since 
I will only examine the relative contributions of metals to the IGM. 
I further assume:

\noindent
\begin{enumerate}
\item The trend of $y_{eff}$ vs. $V_{rot}$ is well described by the function
\begin{equation}
log(y_{eff}) = -1.95 - {(320 - V_{rot})^4\over 9.1\times 10^9}.
\end{equation}
I assume that spirals with $V_{rot}$ $>$ 150 km s$^{-1}$ retain all 
their newly-produced metals with log $y_{eff}$ = --1.95.
The difference between this value and $y_{eff}$($V_{rot}$) given by
the above expression represents the fraction of metals lost by a galaxy.
\noindent
\item The relation between $V_{rot}$ and $M_B$ for my galaxy sample can
be represented by the expression
\begin{equation}
M_B = -6.8~log(V_{rot}) - 4.56.
\end{equation}
This is not meant to be a new derivation of the Tully-Fisher relation,
only a parameterization of the data for this galaxy sample.
\noindent
\item The O/H-luminosity relation for the galaxy sample is represented by 
\begin{equation}
log(O/H) = -0.16 M_B - 6.4.
\end{equation}
\noindent
\item $M/L$ is taken to be equal to one for all galaxies. A factor two
error in $M/L$ will turn out to have only a small effect on the relative
contributions of ejected metals.
\end{enumerate}

As a crude approximation I take the total mass of oxygen in a galaxy 
to be the measured oxygen mass fraction times the mass in stars. Then
the mass of oxygen lost by the galaxy is
\begin{equation}
M_{lost}(O) = 12~(O/H)~L_B~{M\over L}~{0.0112 \over y_{eff}(V_{rot})} - 1,
\end{equation}
where 0.0112 is the average $y_{eff}$ for the massive spirals. The factor
twelve is the conversion from the number ratio O/H to oxygen mass fraction.
Convolving this expression with the galaxy luminosity function gives
the relative mass of oxygen ejected by galaxies of a given luminosity 
into the IGM in a given volume element. The results for this set of
assumptions is illustrated in Figure 7, where I show the relative mass
of oxygen ejected by of all galaxies of a given $M_B$ or $V_{rot}$.

Figure 7 indicates that dwarf galaxies with $M_B$ $>$ --16, $V_{rot}$
$<$ 50 km s$^{-1}$ completely dominate the enrichment of the IGM in
the present day universe. Although the approximations made above are
crude in some cases, this conclusion is fairly robust to those 
approximations. A systematic increase in M/L for more massive galaxies
would shift the distribution to higher $M_B$ and $V_{rot}$, but is not
likely to make the dwarfs less dominant. A steeper luminosity function
would only increase the relative contribution of the dwarf galaxies. 
\cite{trent94} has similarly noted that dwarf galaxies may contribute 
the bulk of the intracluster gas in galaxy clusters, although \cite{gm97}
argue that the dwarf galaxies are unlikely to account for all of the
ICM in clusters. Note that the analysis presented here does not consider
the total amount of gas that may be ejected by galaxies, and that there
are other ways to account for the ICM, such as ram pressure stripping
and tidal stripping of gas ejected during disk galaxy mergers in dense
cluster environments \citep{mihos01}. 

If dwarf galaxies dominate in contributing metals to the IGM, the
abundance pattern of the intergalactic medium should reflect that
of the dwarf galaxies. Irregular galaxies tend to be deficient in
nitrogen and carbon relative to oxygen. Fe/O is not well known in
ionized nebulae because of depletion of iron onto grains, but could
be subsolar or solar, depending on the galaxy's star formation history.
If the outflows consist preferentially of supernova ejecta from massive
stars, then we would expect C, N and Fe to be deficient relative to 
O and other alpha-capture elements. Note that this analysis may not
apply to Ly$\alpha$ systems at high redshift. These systems have 
young ages, and comparison with the Galactic halo abundance pattern
may be more relevant, although in many respects the abundances in
metal-poor dwarf galaxies are similar to those in the halo. Nevertheless,
we might expect a similar relationship between metal loss and rotation
speed to hold in the early universe as well, determining which systems
contribute material to the IGM.

Ejection of metals by galaxies is closely connected to the question of 
feedback of stellar energy into galaxies, a subject of intense interest 
currently. Hierarchical clustering models for galaxy formation have
some difficulty in reproducing the relation between luminosity and
rotation speed in disk galaxies; too much angular momentum is transferred
from the disk to the halo during the collapse, leading to disks that 
rotate too fast for their size. Feedback of energy from stars and 
supernovae into the ambient ISM is expected to relieve this problem,
by reheating the gas and preventing it from collapsing too quickly.
How feedback should be parameterized in galaxy formation models is
poorly understood, however, because star formation itself is poorly
understood. The fraction of metals ejected by a galaxy must be a
function of the feedback process. This connection is complex, however,
and is beyond the scope of this paper. Still, the relation between 
effective yield and rotation speed in galaxies, if it results from the
loss of metals in galactic winds, must be related to the feedback of
energy from stars into the ISM, and thus should be reproduced by 
successful models of feedback in galaxy evolution.


\section{Summary}

I have compiled data on gas fractions and oxygen abundances for a sample 
of approximately 40 spiral and irregular galaxies with rotation speeds
ranging from a few km s$^{-1}$ to 300 km s$^{-1}$. From these data I
have derived effective yields for each galaxy based on the closed box
chemical evolution model; variations in effective yields between galaxies
assess the relative importance of gas flows in the chemical evolution of
those galaxies. I find that the effective yield varies systematically
by a factor of 10-20 as $V_{rot}$ increases from 5 km s$^{-1}$ to 300
km s$^{-1}$, asymptically approaching a constant value for rotation speeds 
greater than 125-150 km s$^{-1}$. This variation appears to be a simple
function of the galaxy potential. If loss of newly-synthesized metals 
in galactic outflows is the main cause of the trend in $y_{eff}$, the 
results suggest that massive spirals essentially retain all of the gas 
ejected in outflows, while dwarf irregulars can lose up to 90\% of their 
metals; the fraction lost is a simple function of the galaxy rotation 
speed. A crude estimate of the mass of metals lost by a given galaxy 
combined with the field galaxy luminosity function indicates that 
galaxies with $M_B$ $>$ --16 and $V_{rot}$ $<$ 50 km s$^{-1}$ should
dominate the contribution of outflows to the enrichment of the IGM.

\acknowledgements

I thank Eric Bell for very many educational discussions about surface
photometry and mass-to-light ratios and a careful reading of the 
manuscript. Robert Braun and Jay Gallagher raised interesting questions 
that made me think harder about the various processes that affect the 
effective yield. Thanks also go to the referee, Mike Edmunds, for 
several very useful comments. This work has made extensive
use of the NASA/IPAC Extragalactic Database (NED), which is operated by
the Jet Propulsion Laboratory, California Institute of Technology, under
contract with the National Aeronautics and Space Administration. Support 
from NASA LTSA grant NAG5-7734 is also acknowledged.

\clearpage

\begin{figure}
\vspace{16.0cm}
\includegraphics{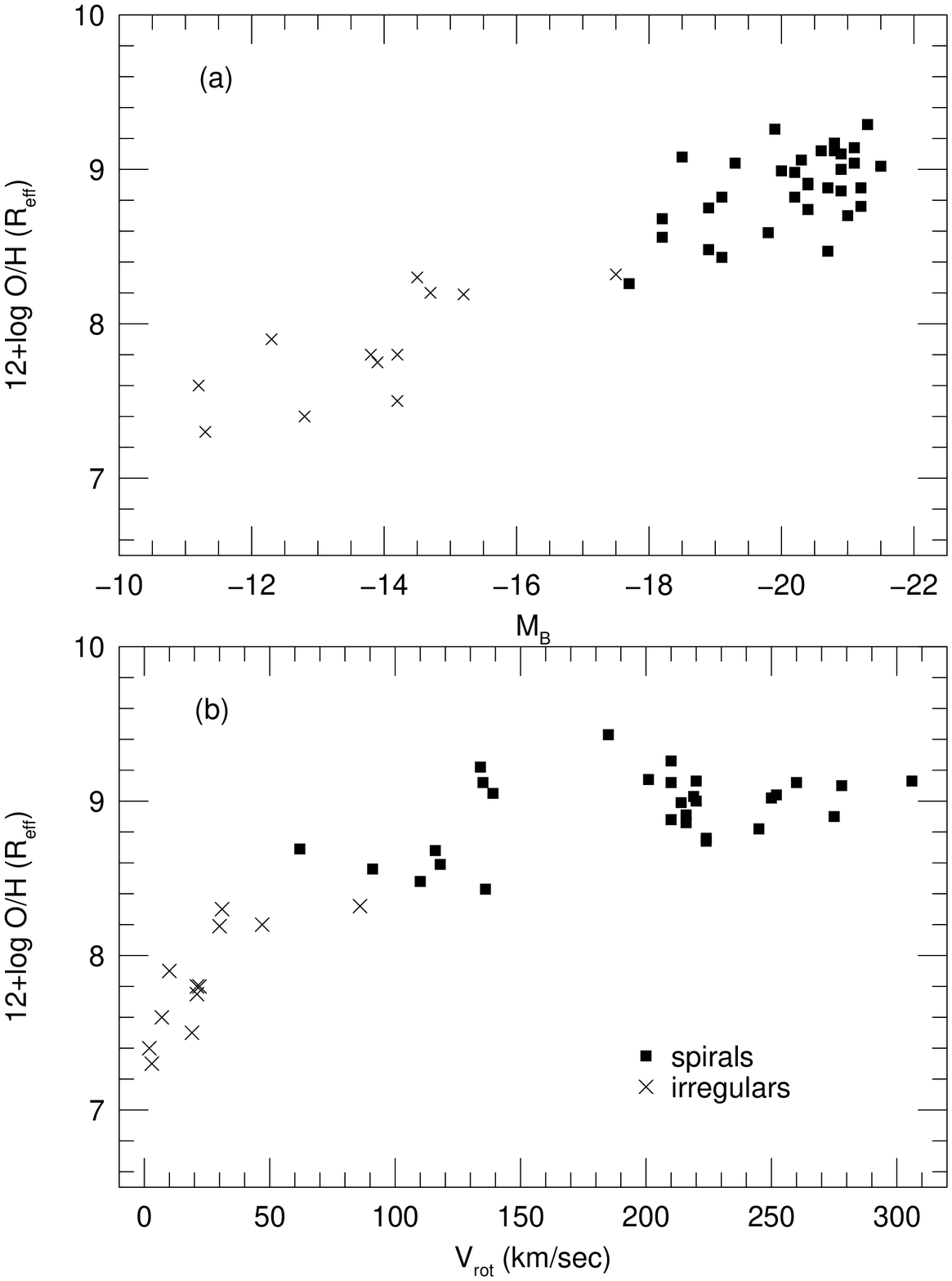}
\figcaption{(a): The correlation between O/H at the disk half-light radius
$R_{eff}$ and absolute B magnitude for spiral and irregular galaxies.
(b) O/H at $R_{eff}$ versus rotation speed $V_{rot}$ for the same sample. For 
spirals the rotation speed is taken to be that on the flat part of the 
rotation curve. 
}
\end{figure}

\clearpage 

\begin{figure}
\vspace{16.0cm}
\includegraphics{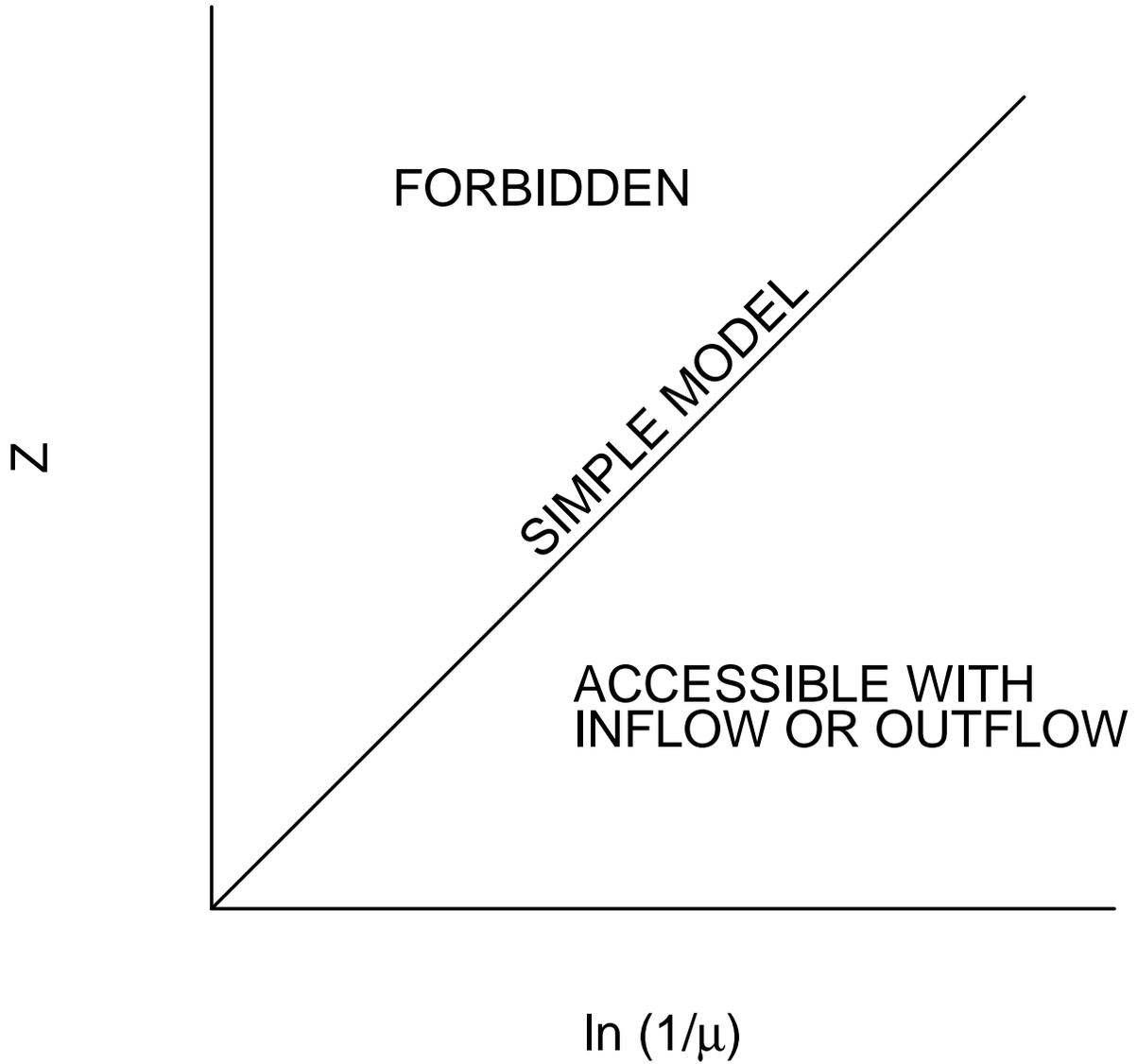}
\caption{Schematic diagram of metallicity Z vs. the logarithm of the 
inverse of the gas fraction $\mu$ for chemical evolution of systems.
The straight line shows the behavior for the simple chemical evolution 
model without gas flows. The region in the upper left can not be attained
by systems with outflow or unenriched inflows of gas. Adapted from 
Figure 1 of \cite{edm90}.
}
\end{figure}

\clearpage 

\begin{figure}
\vspace{16.0cm}
\includegraphics{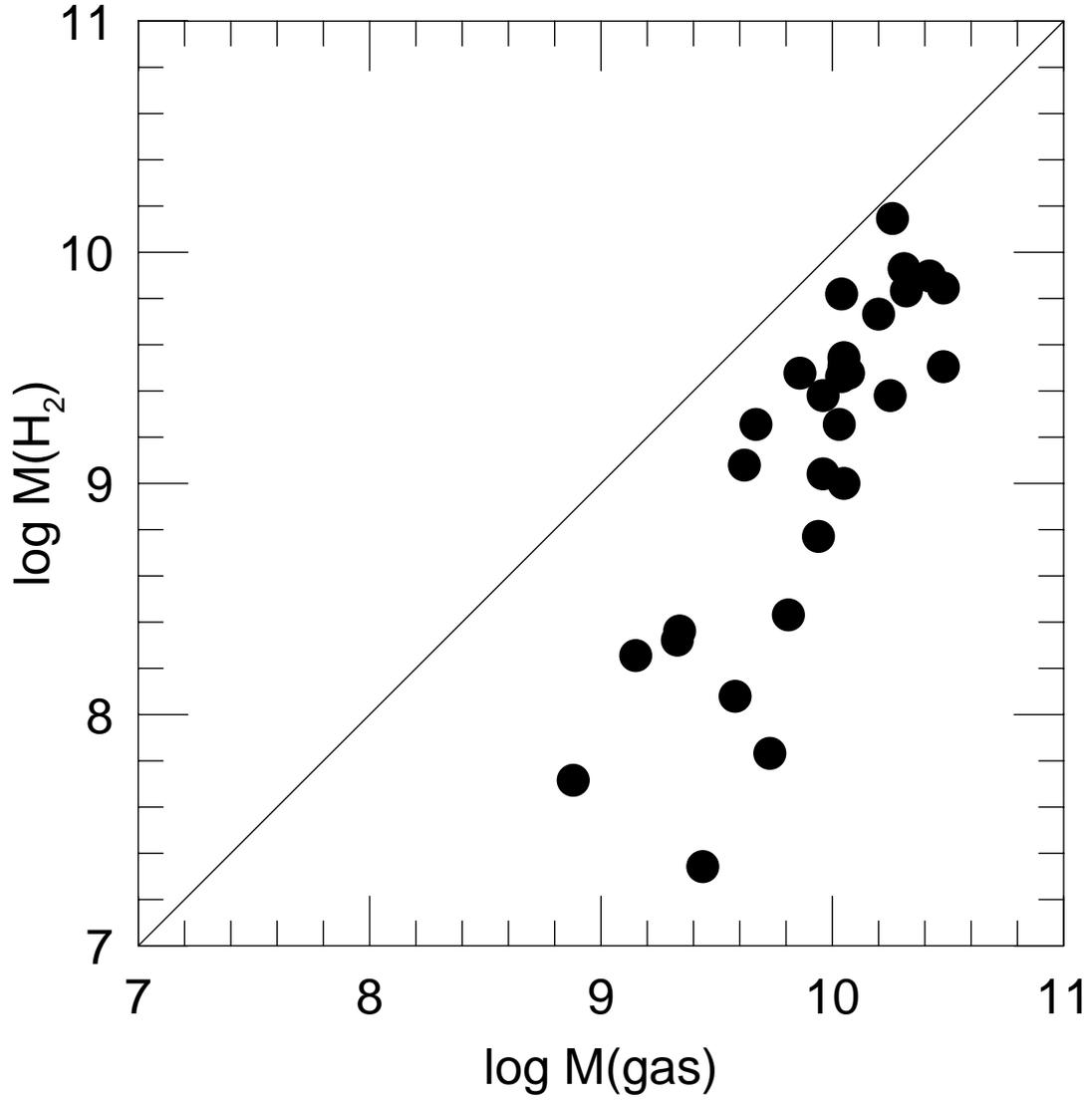}
\caption{Comparison of the mass of molecular gas versus the total gas
mass for the galaxies in Table 1. The solid line denotes equality. See
text for details.
}
\end{figure}

\clearpage 

\begin{figure}
\vspace{16.0cm}
\includegraphics{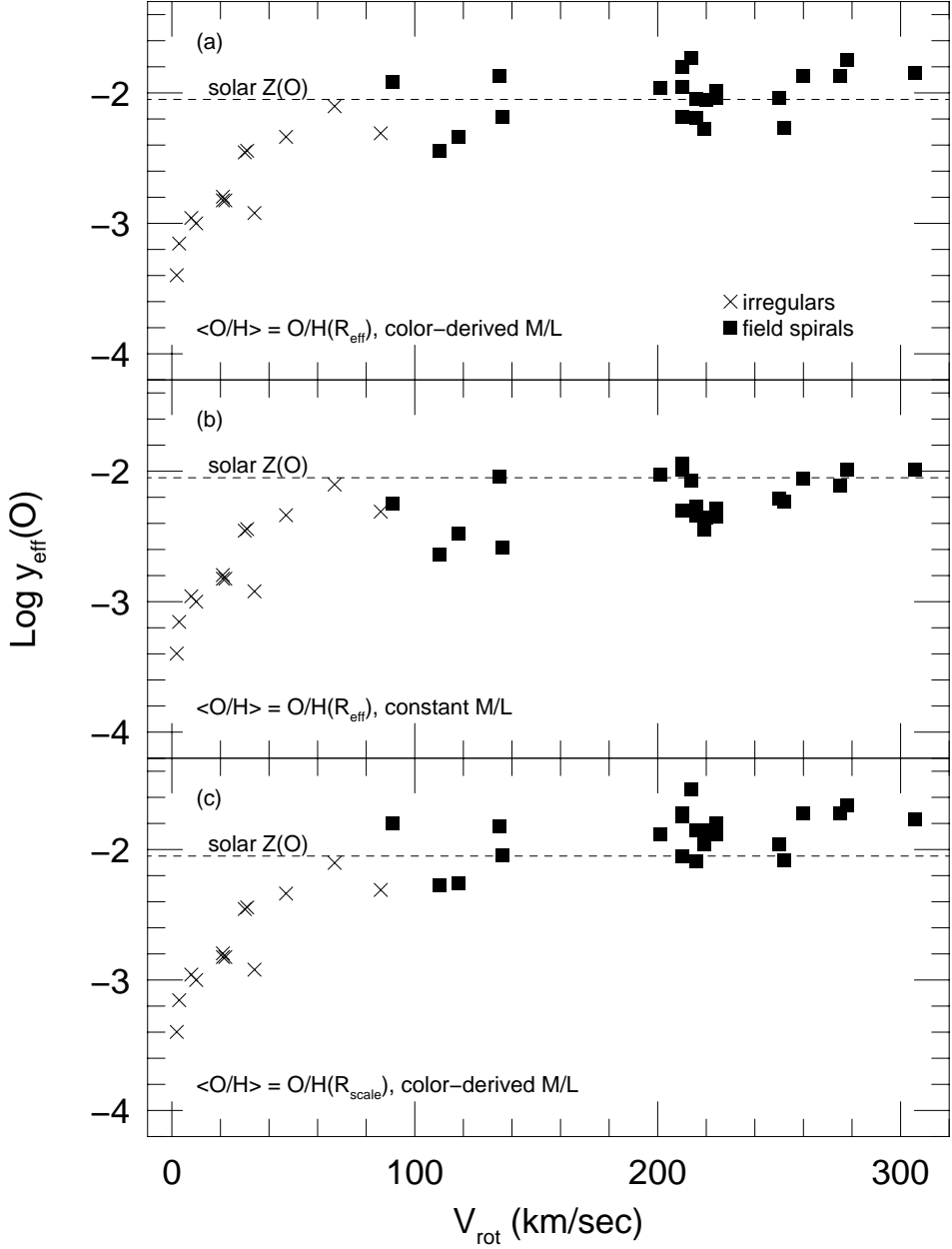}
\figcaption{ The correlation between effective yield $y_{eff}(O)$ and rotation
speed $V_{rot}$ for spiral and irregular galaxies. (a) The average O/H is
assumed to be the value at one disk scale length. Stellar masses derived
using color-dependent M/L ratios based on synthesis models of Bell \& de Jong
(2001). 
(b) Same average O/H as in (a), but using constant M/L = 2 for spiral disks.
(c) Average O/H = O/H at one disk half-light radius; constant M/L as in (b).
The horizontal dashed line shows the value of the solar oxygen mass fraction;
this is for illustration purposes only.
}
\end{figure}

\clearpage 

\begin{figure}
\vspace{16.0cm}
\includegraphics{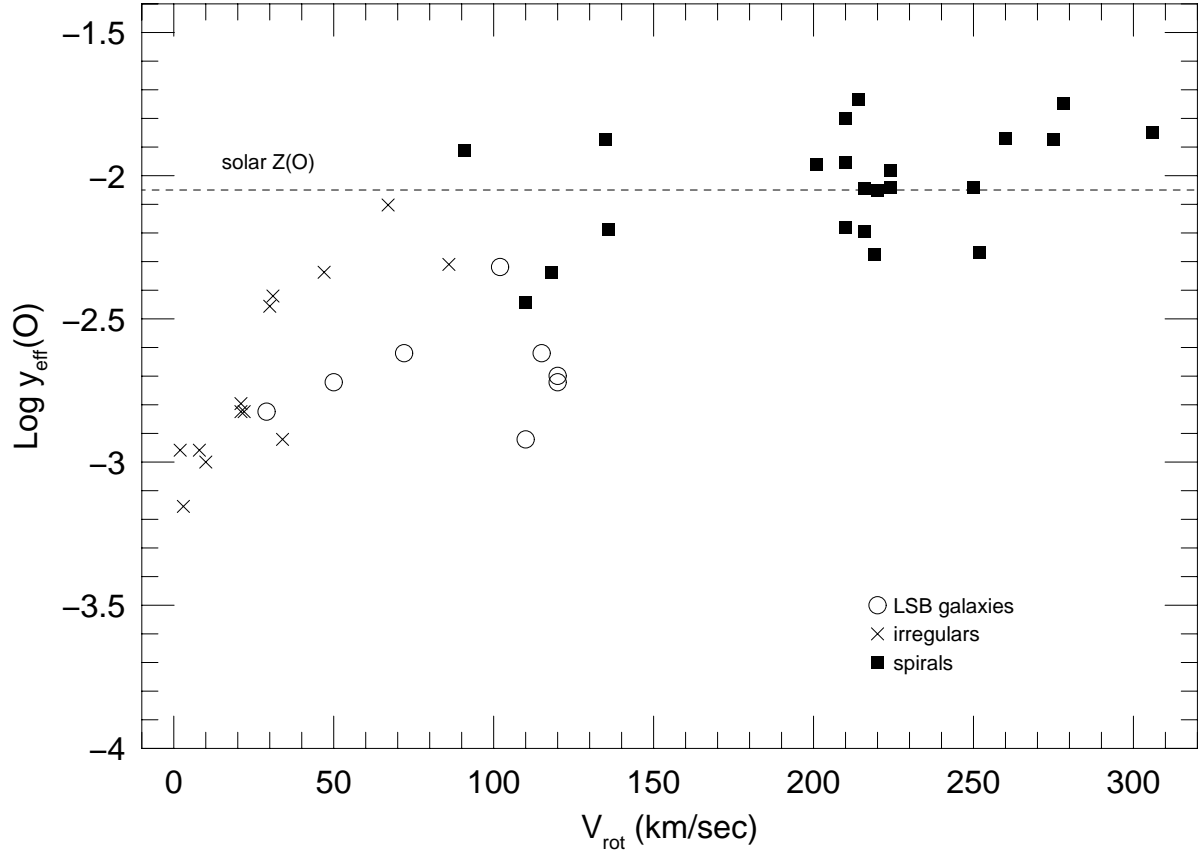}
\caption{As Figure 4(a), but with data for low surface brightness galaxies
included ({\it unfilled circles}). Data for LSB galaxies from \cite{vdhoek00}, 
modified as described in the text. 
}
\end{figure}

\clearpage 

\begin{figure}
\epsscale{0.65}
\vspace{16.0cm}
\includegraphics{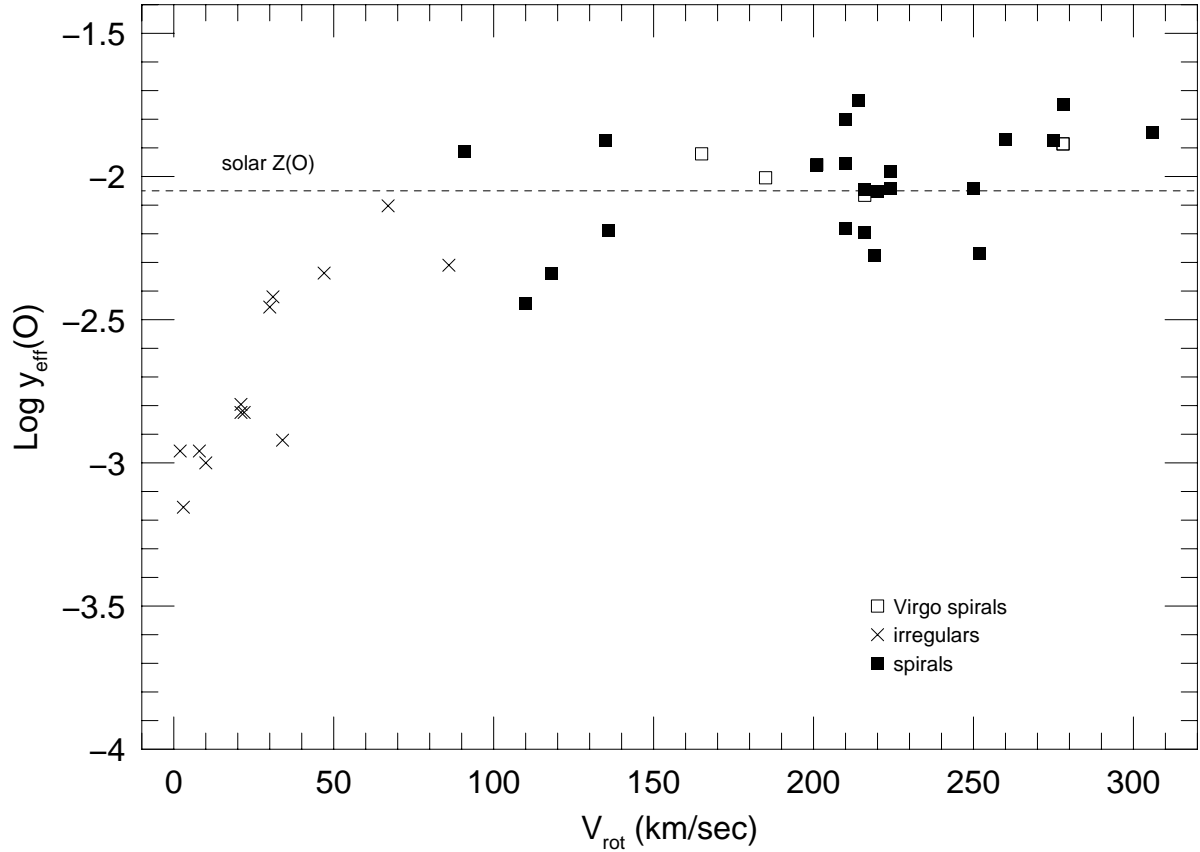}
\caption{As Figure 4(a), but with data for \hi-deficient Virgo spirals 
included \citep{sksz96}. No systematic difference in $y_{eff}$ is seen
between the stripped Virgo spirals and similar field spirals.
}
\end{figure}

\clearpage

\begin{figure}
\vspace{16.0cm}
\includegraphics{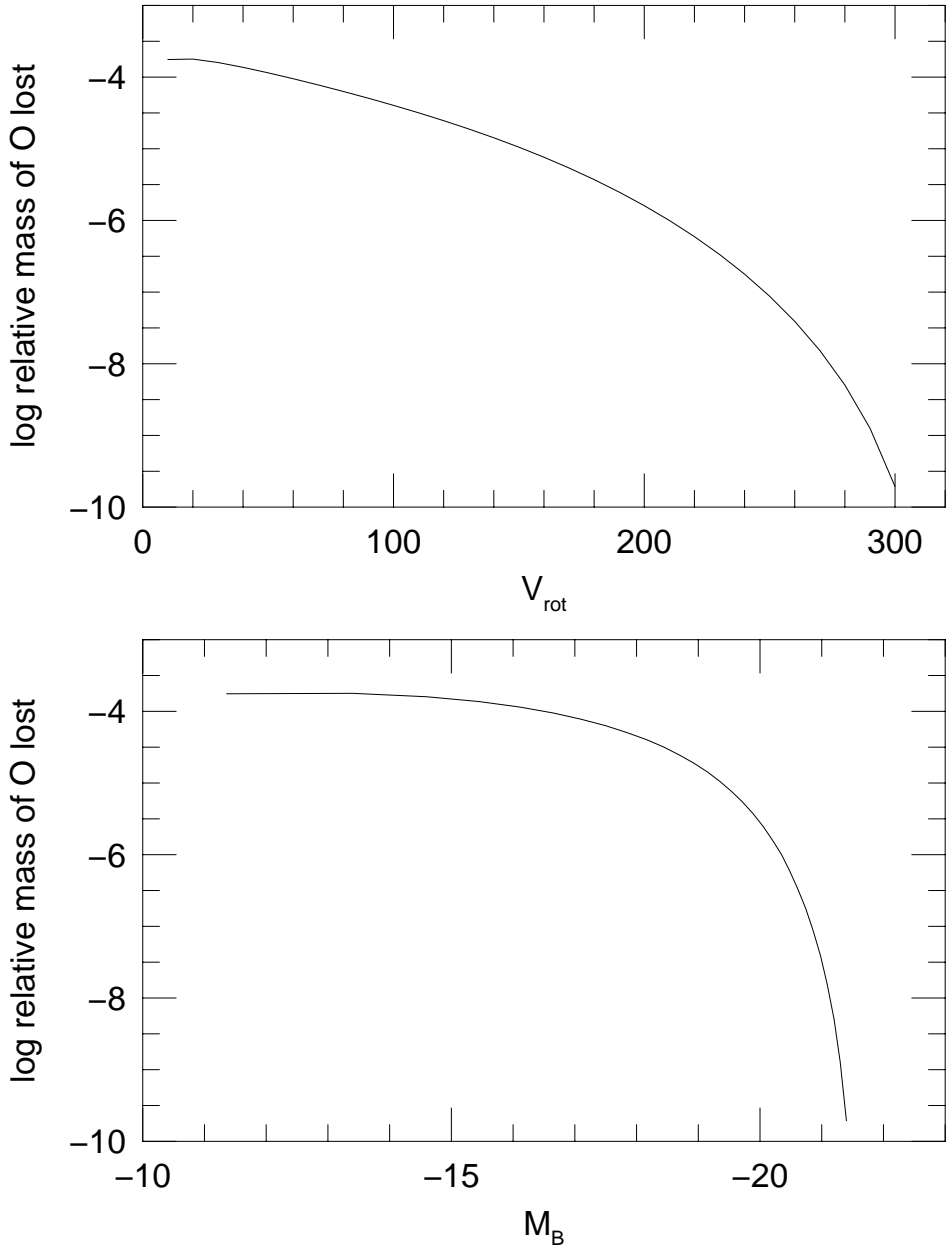}
\figcaption{ The logarithm of the relative contribution of metals to the IGM 
by galaxies as a function of absolute B magnitude $M_B$ ({\it bottom panel}) 
and rotation speed $V_{rot}$ ({\it top panel}). See text for details.
}
\end{figure}


\begin{deluxetable}{lccccccccc}
\tablewidth{0pc}
\tablecaption{Structural Data for Spiral Galaxies}
\tablehead{
\colhead{Galaxy} &\colhead{T} & \colhead{D} & \colhead{Ref.\tablenotemark{a}} & \colhead{M$_B$} &
\colhead{(B--V)$^0_T$} & \colhead{R$_d$} & \colhead{12 + log } & \colhead{Ref.\tablenotemark{b}} &
\colhead{V$_{rot}$} \\ 
\colhead{} & \colhead{} & \colhead{(Mpc)} & \colhead{} & \colhead{} &\colhead{} &
\colhead{(kpc)} & \colhead{O/H($R_{eff}$)} & \colhead{} & \colhead{(km s$^{-1}$)}
}
\startdata
NGC  224 & 3 &  0.77 &  1     & --21.1 & 0.68 &  3.2  &  9.04    & 1,2,3    &   252   \\
NGC  253 & 5 &  2.5  &  2     & --20.0 & \nodata &  3.7  &  8.88    &   4      &   210   \\
NGC  300 & 7 &  2.1  &  3     & --18.1 & 0.58 &  2.1  &  8.56    & 4,5,6    &    91   \\
NGC  598 & 6 &  0.85 &  4     & --18.9 & 0.47 &  2.1  &  8.48    & 7,8,9,10 &   110   \\
NGC  628 & 5 &  9.2  &  5     & --20.3 & 0.51 &  4.0  &  8.90    & 7,11,13  &   275   \\
NGC  925 & 7 &  9.3  &  6     & --19.8 & 0.50 &  4.0  &  8.59    & 12,13    &   118   \\
NGC 1232 & 5 & 21.5  &  7     & --21.3 & 0.59 &  6.3  &  8.88    &  13      &   220   \\
NGC 1637 & 5 &  8.6  &  7     & --18.5 & 0.58 &  0.8  &  9.08    &  13      &   135   \\
NGC 2403 & 6 &  3.2  &  8     & --19.1 & 0.39 &  2.0  &  8.43    & 7,13,14  &   136   \\
NGC 2442 & 4 & 17.1  &  5     & --20.8 & 0.62 &  3.6  &  9.17    &  15      &   134   \\
NGC 2805 & 7 & 23.5  &  7     & --20.8 & 0.44 &  8.5  &  8.47    & 7,13     &   160   \\
NGC 2903 & 4 &  6.3  &  5     & --19.9 & 0.55 &  1.9  &  9.26    & 7,12,13  &   210   \\
NGC 3031 & 2 &  3.6  &  9     & --20.4 & 0.82 &  3.0  &  8.82    & 16,17    &   245   \\
NGC 3344 & 4 &  6.1  &  5     & --18.5 & 0.57 &  1.8  &  8.75    & 7,12     &   165   \\
NGC 3521 & 4 &  7.2  &  5     & --20.0 & 0.68 &  2.3  &  9.12    &  12      &   210   \\
NGC 3621 & 7 &  6.7  & 11     & --20.1 & 0.52 &  2.0  &  8.98    &  15      &   139   \\
NGC 4254 & 5 & 16.1  & (12)   & --21.0 & 0.51 &  3.2  &  9.10    & 7,18     &   250   \\
NGC 4258 & 4 &  7.5  & 13     & --20.8 & 0.55 &  4.4  &  8.91    & 12,19    &   216   \\
NGC 4303 & 4 & 16.1  & (12)   & --21.0 & 0.50 &  4.1  &  8.86    & 18,20    &   150   \\
NGC 4321 & 4 & 16.1  & 12     & --21.1 & 0.65 &  5.2  &  9.14    & 7,18     &   270   \\
NGC 4395 & 9 &  3.6  &  5     & --17.2 & 0.46 &  3.3  &  8.26    & 7,13     &    90   \\
NGC 5033 & 5 & 18.7  &  5     & --21.1 & 0.46 &  6.6  &  8.74    &  12      &   220   \\
NGC 5055 & 4 &  7.2  &  5     & --20.2 & 0.64 &  4.0  &  8.99    & 7,4      &   214   \\
NGC 5194 & 4 &  7.7  &  5,(14) & --20.8 & 0.53 &  4.3  &  9.12   & 7,21    &   260   \\
NGC 5236 & 5 &  3.6  & (15),(16) & --19.8 & 0.61 &  2.8  &  9.06 &   4,22   &   180   \\
NGC 5457 & 6 &  7.2  & 14     & --21.1 & 0.44 &  5.4  &  8.76    &   23     &   260   \\
NGC 6384 & 4 & 26.6  &  5     & --21.5 & 0.55 & 11.8  &  8.70    &  19      &   219   \\
NGC 6744 & 4 & 10.4  &  5     & --21.3 & 0.61 &  5.0  &  9.29    &  15      &   200   \\
NGC 6946 & 6 &  5.5  &  5     & --20.9 & 0.40 &  2.7  &  8.87    & 11       &   220   \\
NGC 7331 & 4 & 15.1  & 17     & --21.5 & 0.63 &  4.8  &  9.02    & 12,19    &   250   \\
NGC 7793 & 8 &  2.8  &  5     & --17.9 & 0.51 &  1.4  &  8.68    &  4,7     &   116   \\
\enddata
\tablenotetext{a}{References for distances (parentheses around column entry 
denote association with galaxy having measured distance): (1) Freedman \& Madore 1990;
(2) Miller 1996; (3) Freedman et al. 1992; (4) Freedman, Wilson, \& Madore 1991;
(5) Tully 1988; (6) Silbermann et al. 1996; (7) van Zee et al. 1998; (8) Freedman
\& Madore 1988; (9) Freedman et al. 1994; (10) Kelson et al. 1999 (NGC 3198); (11)
Rawson et al. 1997; (12) Ferrarese et al. 1996; (13) Newman et al. 2001; (14)
Kelson et al. 1996; (15) Saha et al. 1995; (16) Soria et al. 1996; (17) Hughes 
et al. 1998}
\tablenotetext{b}{References for oxygen abundances: (1) Dennefeld \& Kunth 1981;
(2) Blair, Kirshner, \& Chevalier 1982; (3) Galarza, Walterbos, \& Braun 1999; 
(4) Webster \& Smith 1983; (5) Pagel et al. 1979; (6) Deharveng et al. 1988; 
(7) McCall, Rybski, \& Shields 1985; (8) Kwitter \& Aller 1981; (9) V\'\i lchez 
et al. 1988; (10) Smith 1975; (11) Ferguson, Gallagher, \& Wyse 1998; (12) 
Zaritsky, Kennicutt, \& Huchra 1994; (13) van Zee et al. 1998; (14) Garnett et 
al. 1997; (15) Ryder 1995; (16) Garnett \& Shields 1987; (17) Stauffer \&
Bothun 1984; (18) Skillman et al. 1996; (19) Oey \& Kennicutt 1993; (20) 
Henry et al. 1992; (21) D\'\i az et al. 1991; (22) Dufour et al. 1980; 
(23) Kennicutt \& Garnett 1996 }
\end{deluxetable}

\clearpage

\begin{deluxetable}{lccccccc}
\tablewidth{40pc}
\tablecaption{Gas Data for Spiral Galaxies}
\tablehead{
\colhead{Galaxy} & \colhead{} & \colhead{log M(H~I)} & \colhead{} & \colhead{Reference\tablenotemark{a}} & 
\colhead{log M(H$_2$)} & \colhead{Reference\tablenotemark{b}} \\
\colhead{} & \colhead{} & \colhead{(M$_\odot$)} & \colhead{} & \colhead{} & \colhead{M$_\odot$} & \colhead{} }
\startdata
NGC  224 & &  9.68 & & 1,2,3,4,5              &   8.45  & 1        & \\
NGC  253 & &  9.34 & &  6,7,8                 &   9.28  & 2,3      & \\
NGC  300 & &  9.41 & &  9                     & \nodata &          & \\
NGC  598 & &  9.32 & & 4,10,11,12             &   7.35  & 4        & \\
NGC  628 & & 10.07 & & 4,13,14,15,16,17,18,19 &   9.38  & 5,6      & \\
NGC  925 & &  9.79 & & 4,15,16,17,20,21,22,23 &   8.78  & 5        & \\
NGC 1232 & & 10.14 & & 19,21,24,25,26,27      & \nodata &          & \\
NGC 1637 & &  9.18 & & 16,28,29,30            &   8.34  & 7        & \\
NGC 2403 & &  9.61 & & 13,17,31,32            &   7.84  & 8        & \\
NGC 2442 & &  9.78 & & 26,33                  &   9.51  & 9        & \\
NGC 2805 & & 10.13 & & 16,21,23,24,34,35,36   & \nodata &          & \\
NGC 2903 & &  9.36 & & 19,29,37,38,39         &   9.11  & 5,6,10   & \\
NGC 3031 & &  9.45 & & 15,40,41,42,43         &   8.28  & 5,11     & \\
NGC 3344 & &  9.18 & & 15,17,22,23,29,44.45   &   8.37  & 6        & \\
NGC 3521 & &  9.52 & & 15,17,21,23,45         &   9.49  & 6        & \\
NGC 3621 & &  9.97 & & 8,26                   & \nodata &          & \\
NGC 4254 & &  9.97 & & 46                     &   9.94  & 12,13    & \\
NGC 4258 & &  9.78 & & 17,21,37,45,47,48      &   9.06  & 6,14     & \\
NGC 4303 & &  9.90 & & 16,46,49               &   9.75  & 13,15,16 & \\
NGC 4321 & &  9.49 & & 16,46,50               &  10.16  & 15,17    & \\
NGC 4395 & &  8.97 & & 17,21,30,37            &   8.28  & 18       & \\
NGC 5033 & & 10.26 & & 17,21,29               &   9.86  & 6,18     & \\
NGC 5055 & &  9.75 & & 17,45                  &   9.55  & 5,6      & \\
NGC 5194 & &  9.53 & & 13,17,51,52            &   9.88  & 6,19     & \\
NGC 5236 & &  9.72 & & 4,53,54,55             &   9.50  & 6        & \\
NGC 5457 & & 10.32 & & 4,17,56,57,58,59       &   9.51  & 20       & \\
NGC 6384 & & 10.15 & & 16,24,30,60            &   9.91  & 6        & \\
NGC 6744 & & 10.40 & & 26                     & \nodata &          & \\
NGC 6946 & &  9.80 & & 4,17,61                &   9.47  & 21,22    & \\
NGC 7331 & & 10.03 & & 16,17,21,30,45,62      &   9.84  & 6,23     & \\
NGC 7793 & &  8.73 & &  8,15,21,26,63         &   7.72  & 18       & \\
\enddata
\vfill\eject
\tablenotetext{a}{References for H~I data: (1) Heideman 1961; (2) Argyle 1965;
(3) Gottesman \& Davies 1970; (4) Dean \& Davies 1975; (5) Cram, Roberts, \&
Whitehurst 1980; (6) Huchtmeier 1972; (7) Combes, Gottesman, \& Weliachew 1977;
(8) Whiteoak \& Gardner 1977; (9) Shobbrook \& Robinson 1967; (10) Dieter 1962;
(11) Gordon 1971; (12) Huchtmeier 1972b; (13) Roberts 1962; (14) Roberts 1969;
(15) Gougenheim 1969; (16) Shostak 1978; (17) Rots 1980; (18) Briggs 1982: 
(19) Staveley-Smith \& Davies 1988; (20) H\"oglund \& Roberts 1966; (21)
Fisher \& Tully 1981; (22) Hewitt, Haynes \& Giovanelli 1983; (23) Davis
\& Seaquist 1983; (24) Bottinelli, Gougenheim, \& Paturel 1982; (25) Becker
et al. 1988; (26) Reif et al. 1982; (27) van Zee \& Bryant 1999; (28)
Bottinelli et al. 1970; (29) Roberts 1968; (30) Haynes et al. 1998; (31)
Burns \& Roberts 1971; (32) Shostak \& Rogstad 1973; (33) Bajaja \& Martin
1985; (34) Bosma et al. 1980; (35) Reakes 1979; (36) Dickel \& Rood 1980; 
(37) Huchtmeier \& Seiradakis 1985; (38) Huchtmeier \& Richter 1986; (39)
Wevers, van der Kruit, \& Allen 1986; (40) Rots \& Shane 1974; (41) Rots
\& Shane 1975; (42) Appleton, Davies \& Stephenson 1981; (43) Yun, Ho, \&
Lo 1994; (44) Lewis \& Davies 1987; (45) Staveley-Smith \& Davies (1987);
(46) Davies \& Lewis 1973; (47) van Albada 1980; (48) Appleton \& Davies
1982; (49) McCutcheon \& Davies 1970; (50) Gallagher, Faber, \& Balick 1975;  
(51) Roberts \& Warren 1970; (52) Tilanus \& Allen 1991; (53) Bottinelli
\& Gougenheim 1973; (54) Huchtmeier \& Bohnenstengel 1981; (55) Epstein 1964;
(56) Huchtmeier \& Witzel 1979; (57) Gu\'elin \& Weliachew 1970; (58) Bosma,
Goss, \& Allen 1981; (59) Davies, Davidson, \& Johnson 1980; (60) Pfleiderer
et al. 1981; (61) Gordon, Remage, \& Roberts 1968; (62) Shostak \& Allen 1980;
(63) Carignan 1985}
\tablenotetext{b}{References for CO data: (1) Koper et al. 1991; (2) Sorai et al.
2000; (3) Houghton et al. 1997; (4) Wilson \& Scoville 1989; (5) Sage 1993;
(6) Young et al. 1995; (7) Adler \& Liszt 1989; (8) Thornley \& Wilson 1995;
(9) Bajaja et al. 1995; (10) Jackson et al. 1989; (11) Brouillet et al. 1991;
(12) Knapp, Helou, \& Stark 1987; (13) Kenney \& Young 1989; (14) Cox \&
Downes 1996; (15) Stark et al. 1986; (16) Tinney et al. 1990; (17) Sempere
\& Garc\'\i a-Burillo 1997; (18) Stark, Elmegreen \& Chance 1987; (19)
Kuno et al. 1995; (20) Kenney, Scoville, \& Wilson 1991; (21) Young \& Scoville
1982a; (22) Tacconi \& Young 1989; (23) Young \& Scoville 1982b
}
\end{deluxetable}

\clearpage

\begin{deluxetable}{lcccccccccc}
\tablewidth{0pc}
\tablecaption{Data for Irregular Galaxies}
\tablehead{
\colhead{Galaxy} &\colhead{T} & \colhead{D} & \colhead{Ref.\tablenotemark{a}} & \colhead{M$_B$} &
\colhead{B--V} & \colhead{log M$_{gas}$\tablenotemark{b}} & \colhead{Ref.\tablenotemark{c}} & \colhead{12 + log} & \colhead{Ref.\tablenotemark{d}} &
\colhead{V$_{rot}$} \\ 
\colhead{} & \colhead{} & \colhead{(Mpc)} & \colhead{} & \colhead{} &\colhead{} &
\colhead{(M$_{\odot}$)} & \colhead{} & \colhead{O/H} & \colhead{} & \colhead{(km s$^{-1}$)}
}
\startdata
WLM       & 10 & 0.92 &  1  & --13.9 & 0.60 &  7.90 &  1  & 7.75 & 1,2 & 21  \\
NGC 55    &  9 & 1.5\ & 2,3 & --17.5 & 0.47 &  9.27 &  2  & 8.32 & 3,4 & 86  \\
IC 10     &  9 & 0.82 &  4  & --15.2 & 0.50 &  8.30 &  1  & 8.19 &  5  & 30  \\
IC 1613   &  9 & 0.70 & 5,6 & --14.2 & 0.57 &  7.97 & 3,4 & 7.8\ &  3  & 21  \\
Leo A     & 10 & 0.69 &  7  & --11.3 & 0.15 &  7.00 &  5  & 7.3\ &  6  &  3  \\
Sextans B & 10 & 1.34 &  8  & --13.8 & 0.47 &  7.80 & 1,4 & 7.8\ & 1,7 & 22  \\
Sextans A & 10 & 1.44 &  9  & --14.2 & 0.35 &  8.20 & 6,7 & 7.5\ &  6  & 34  \\
DDO 155   & 10 & 1.6\ & 10  & --11.2 & 0.35 &  6.84 & 4,5 & 7.6\ &  7  &  8  \\
Sgr dI    & 10 & 1.2\ & 11  & --11.6 & 0.41 &  7.11 & 8,9 & 7.4\ &  1  &  2  \\
NGC 6822  &  9 & 0.49 &  1  & --14.7 & 0.47 &  8.27 &  1  & 8.2\ &  8  & 47  \\
IC 5152   & 10 & 1.7\ & 12  & --14.8 & 0.33 &  7.89 &  1  & 8.3\ &  3  & 31  \\
Pegasus   & 10 & 0.96 & 13  & --12.3 & 0.59 &  6.82 &  4  & 7.9\ &  9  & 10  \\
NGC 3109  & 10 & 1.25 & 14  & --15.2 & 0.48 &  8.86 &  1  & 8.06 &  3  & 67  \\
\enddata
\tablenotetext{a}{References for distances: (1) Gallart, Aparicio,
\& V\'\i lchez 1996; (2) Puche, Carignan, \& Wainscoat 1991; (3) Pritchet et al. 
1988; (4) Saha et al. 1996; (5) Saha et al. 1992; (6) Lee et al. 1993; 
(7) Tolstoy et al. 1998; (8) Piotto et al. 1994; (9) Sakai et al. 1996; (10) 
Dohm-Palmer et al. 1998; (11) Lee \& Kim 2000; (12) Zijlstra \& Minniti 1999; 
(13) Aparicio, Gallart, \& Bertelli 1997; (14) Lee 1993. }
\tablenotetext{c} {$M_{gas}$ = 1.33 $M(H~I)$, to correct for helium.}
\tablenotetext{c}{References for H I measurements: (1) Huchtmeier \& Richter 1986;
(2) Robinson \& van Damme 1966; (3) Lake \& Skillman 1989; (4) Hoffman et al. 
1996; (5) Fisher \& Tully 1975; (6) Huchtmeier, Seiradakis, \& Materne 1981; 
(7) Skillman et al. 1988; (8) Longmore et al. 1982; (9) Young \& Lo 1997. }
\tablenotetext{d}{References for oxygen abundances: (1) Skillman et al. 1989a;
(2) Hodge \& Miller 1995; (3) Talent 1980; (4) Webster \& Smith 1983; 
(5) Lequeux et al. 1979; (6) Skillman et al. 1989b; (7) Moles et al. 1990; 
(8) Pagel et al. 1980; (9) Skillman et al. 1997.  }
\end{deluxetable}

\clearpage
\begin{deluxetable}{lcccc}
\tablewidth{0pc}
\tablecaption{Derived Gas Fractions and Effective Yields}
\tablehead{
\colhead{Galaxy} &\colhead{12 + log(O/H)} & \colhead{V$_{rot}$} 
& \colhead{M$_{gas}$/M$_{tot}$} & \colhead{y$_{eff}$\tablenotemark{a}} \\ 
\colhead{} & \colhead{at R$_{eff}$} & \colhead{(km s$^{-1}$)} &
\colhead{} & \colhead{} 
}
\startdata
NGC  224 &  9.04  & 252 &  0.11 & 0.0068   \\
NGC  253 &  8.88  & 210 &  0.16 & 0.0067   \\
NGC  300 &  8.56  &  91 &  0.46 & 0.0076   \\
NGC  598 &  8.48  & 110 &  0.20 & 0.0033   \\
NGC  628 &  8.90  & 275 &  0.31 & 0.011    \\
NGC  925 &  8.59  & 118 &  0.25 & 0.0040   \\
NGC 1637 &  9.08  & 135 &  0.22 & 0.010    \\
NGC 2403 &  8.43  & 136 &  0.29 & 0.0036   \\
NGC 2903 &  9.26  & 210 &  0.14 & 0.014    \\
NGC 3344 &  8.75  & 165 &  0.22 & 0.0074   \\
NGC 3521 &  9.12  & 210 &  0.21 & 0.017    \\
NGC 4254 &  9.10  & 278 &  0.24 & 0.013    \\
NGC 4258 &  8.91  & 216 &  0.12 & 0.0058   \\
NGC 4303 &  8.86  & 216 &  0.20 & 0.0086   \\
NGC 4321 &  9.14  & 201 &  0.17 & 0.011    \\
NGC 5033 &  8.74  & 224 &  0.28 & 0.0078   \\
NGC 5055 &  8.99  & 214 &  0.25 & 0.013    \\
NGC 5194 &  9.12  & 260 &  0.16 & 0.012    \\
NGC 5236 &  9.06  & 306 &  0.27 & 0.012    \\
NGC 5457 &  8.76  & 224 &  0.22 & 0.0064   \\
NGC 6384 &  8.70  & 219 &  0.19 & 0.0077   \\
NGC 6946 &  8.87  & 220 &  0.13 & 0.0060   \\
NGC 7331 &  9.02  & 250 &  0.13 & 0.0074   \\
WLM      &  7.75  &  21 &  0.46 & 0.0009  \\
NGC 55   &  8.32  &  86 &  0.55 & 0.0042  \\
IC 10    &  8.19  &  30 &  0.49 & 0.0026  \\
IC 1613  &  7.8   &  21 &  0.40 & 0.0008  \\
Leo A    &  7.3   &   3 &  0.89 & 0.0020  \\
Sextans B    &  7.8   &  22 &  0.57 & 0.0013  \\
Sextans A    &  7.5   &  19 &  0.70 & 0.0011  \\
DDO 155  &  7.6   &   7 &  0.71 & 0.0014  \\
Sgr dI   &  7.4   &   2 &  0.69 & 0.0008  \\
NGC 6822 &  8.2   &  47 &  0.60 & 0.0037  \\
IC 5152  &  8.3   &  31 &  0.60 & 0.0047  \\
Pegasus  &  7.9   &  10 &  0.34 & 0.0009  \\
NGC 3109 &  8.06  &  67 &  0.83 & 0.0074  \\
\enddata
\tablenotetext{a}{Effective yield for O assuming average O/H 
is O/H at R$_{eff}$. }
\end{deluxetable}

\end{document}